\documentclass[twocolumn,10pt, amssymb, amsmath, aps, showpacs, footinbib, prb]{revtex4-1}
\usepackage{graphicx}

\newcommand{\eff}{\text{eff}}
\newcommand{\AFM}{\text{AFM}}

\begin{document}

\title{Microscopic model of (CuCl)LaNb$_2$O$_7$:\\ coupled spin dimers replace a frustrated square lattice}

\author{Alexander A. Tsirlin}
\email{altsirlin@gmail.com}
\author{Helge Rosner}
\email{Helge.Rosner@cpfs.mpg.de}
\affiliation{Max Planck Institute for Chemical Physics of Solids, N\"{o}thnitzer
Str. 40, 01187 Dresden, Germany}


\begin{abstract}
We present a microscopic model of the spin-gap quantum magnet (CuCl)LaNb$_2$O$_7$, previously suggested as a realization of the spin-$\frac12$ frustrated square lattice. Taking advantage of the precise atomic positions from recent crystal structure refinement, we evaluate individual exchange integrals and construct a minimum model that naturally explains all the available experimental data. Surprisingly, the deviation from tetragonal symmetry leads to the formation of spin dimers between fourth neighbors due to a Cu--Cl--Cl--Cu pathway with an antiferromagnetic exchange $J_4\simeq 25$~K. The total interdimer exchange amounts to $12-15$~K. Our model is in agreement with inelastic neutron scattering results and is further confirmed by quantum Monte Carlo simulations of the magnetic susceptibility and the high-field magnetization. We establish (CuCl)LaNb$_2$O$_7$ as a non-frustrated system of coupled spin dimers with predominant antiferromagnetic interactions and provide a general perspective for related materials with unusual low-temperature magnetic properties.
\end{abstract}

\maketitle

The spin liquid ground state is one of the great challenges in condensed matter physics.\cite{balents2010,lee2008} While all spin liquids share the absence of the long-range ordering down to zero temperature, the details of their magnetic behavior depend on the specific type of spin correlations, largely determined by the lattice topology. A spin liquid state is readily achieved in many \mbox{spin-$\frac12$} gapped models (spin dimer, alternating chain, two-leg ladder), where the gap in the excitation spectrum results from a singlet ground state without long-range ordering. Two-dimensional (2D) frustrated spin systems show more exotic spin-liquid regimes,\cite{balents2010} but only a limited range of model materials has been studied so far.\cite{PHCC,coldea2001}

The (CuCl)LaNb$_2$O$_7$ compound is commonly referred as an experimental realization of the spin-$\frac12$ frustrated square lattice (FSL) model. This model entails competing nearest-neighbor ($J_1$) and next-nearest-neighbor ($J_2$) couplings on the square lattice and shows the spin liquid ground state in a narrow range of parameters ($J_2/J_1\simeq 0.5$).\cite{misguich} The initially proposed tetragonal symmetry of (CuCl)LaNb$_2$O$_7$ exhibits the square-lattice arrangement of \mbox{spin-$\frac12$} Cu$^{+2}$ cations.\cite{koden1999} The spin-gap behavior gave rise to a common belief that (CuCl)LaNb$_2$O$_7$ is the first experimental example of a spin liquid regime within the FSL model. However, the magnetization data fit poorly to theoretical predictions.\cite{kageyama2005,kageyama2005-2} Moreover, an inelastic neutron scattering experiment evidenced an unusual position of the gap excitation that could be formally assigned to a dimer with the length of 8.8~\r A (compare to 3.8~\r A and 5.5~\r A for $J_1$ and $J_2$, respectively).\cite{kageyama2005} Nuclear magnetic resonance experiments clearly showed a lack of tetragonal symmetry, thus invalidating the FSL-based description.\cite{yoshida2007} Until recently, several competing proposals for the crystal structure and the magnetic model were available,\cite{yoshida2007,tsirlin2009,cheng} but neither of them could fully explain the spin-gap behavior and provide a quantitative interpretation of the experimental data.

Our recent structure refinement, based on a high-resolution x-ray diffraction experiment,\cite{refinement} yielded accurate and reliable atomic positions for (CuCl)LaNb$_2$O$_7$. Computational approaches\cite{cheng,refinement} seem to converge to the same structural model, although the calculated interatomic distances and angles are slightly different due to the inevitable shortcomings of density functional theory (DFT), especially in strongly correlated electronic systems. In the following, we will use the accurate \emph{experimental} structural information to derive individual exchange couplings, to establish the microscopic model, and to resolve the long-standing puzzle of (CuCl)LaNb$_2$O$_7$. We also consider the transferability of our model to closely related materials\cite{oba2006,kitada2009} that reveal similar interpretation problems, although their magnetic behavior is strikingly different.

The evaluation of individual exchange couplings is based on scalar-relativistic DFT band structure calculations within the local density approximation (LDA)\cite{perdew-wang} and local spin density approximation (LSDA)+$U$ approaches. We used the \texttt{FPLO} code with the basis set of atomic-like local orbitals.\cite{fplo} The on-site Coulomb repulsion parameter $U_{3d}$ was varied in a range of $3.5-9.5$~eV, while the exchange parameter $J_{3d}$ was fixed at 1~eV. 

The exchange couplings are calculated via two complementary procedures: 

i) (model approach) The LDA band structure is mapped onto a tight-binding (TB) model and further onto a Hubbard model in the strongly correlated regime $t_i\ll U_{\eff}$, where $t_i$ is a hopping of the TB model and $U_{\eff}$ is the effective on-site Coulomb repulsion in the Cu $3d$ bands (in general, different from $U_{3d}$ applied to the atomic $3d$ orbitals). In the half-filling regime, the low-lying excitations are described by the Heisenberg model with antiferromagnetic (AFM) exchange $J_i^{\AFM}=4t_i^2/U_{\eff}$. This approach evaluates \emph{all} the exchange couplings in the system, yet it does not account for the ferromagnetic (FM) part of the exchange.

ii) (supercell approach) Total energies for a set of ordered spin configurations from LSDA+$U$ are mapped onto a classical Heisenberg model, thus yielding the total exchanges~$J_i$. 

\begin{figure}
\includegraphics{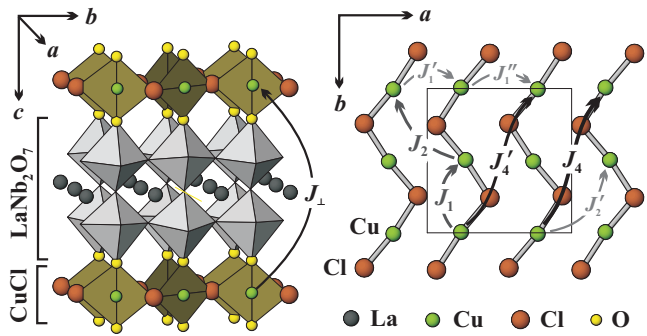}
\caption{\label{structure}
(Color online) Crystal structure of (CuCl)LaNb$_2$O$_7$: the overall view (left panel) and the [CuCl] ``layers'' in the $ab$ plane (right panel). The couplings are labeled according to the Cu--Cu distances: the subscript 1 denotes the interactions between first neighbors, the subscript 2 denotes the interactions between second neighbors, etc.
}
\end{figure}

\begin{figure}
\includegraphics[scale=0.92]{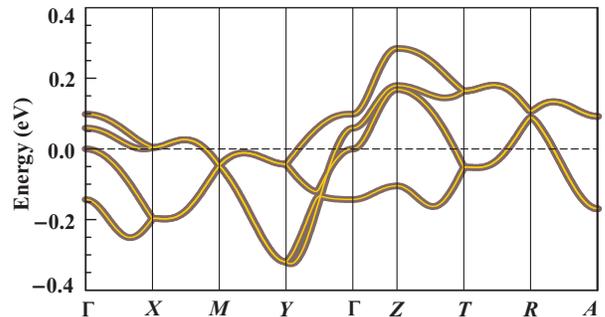}
\caption{\label{tb}
(Color online) Tight-binding fit (thick dark lines) of the LDA band structure (thin light lines).
}
\end{figure}

The crystal structure of (CuCl)LaNb$_2$O$_7$ is shown in Fig.~\ref{structure}. The [CuCl] layers in the $ab$ plane were initially described within a four-fold symmetry that would lead to a square lattice of the Cu$^{+2}$ cations. However, the precise structure determination splits these ``layers'' into chains of corner-sharing CuO$_2$Cl$_2$ plaquettes.\cite{refinement} The LDA valence band structure\cite{supplement} is typical for cuprates. Setting the Fermi level to zero energy, we find: i) the fully filled valence bands below $-0.9$~eV; ii) the half-filled Cu $3d_{x^2-y^2}$ bands between $-0.3$~eV and 0.3~eV (Fig.~\ref{tb}); iii) the wide Nb $4d$ bands above 0.5~eV. The apparent metallicity is a well-known shortcoming of LDA, when applied to strongly correlated systems. The LSDA+$U$ calculations provide a correction to the missing correlation energy in a mean-field approximation and lead to an insulating energy spectrum.\cite{supplement}

\begin{figure}
\includegraphics{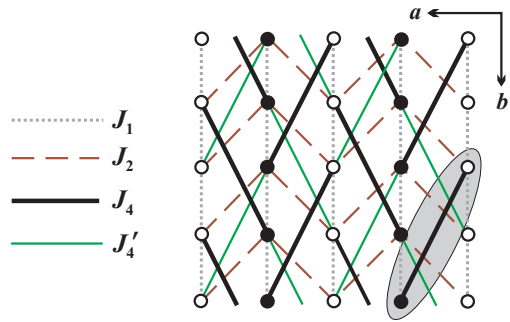}
\caption{\label{model}
(Color online) A sketch of the (CuCl)LaNb$_2$O$_7$ spin lattice in the $ab$ plane. The FM coupling $J_1$ is denoted by the dotted line, the AFM couplings $J_2$, $J_4$, and $J_4'$ are shown by the dashed, thick solid, and thin solid lines, respectively. The couplings $J_{\perp}$ run along the $c$ axis and connect the neighboring planes. The shading denotes the spin dimer. The open and filled circles show the columnar AFM ordering in the case of weak dimerization.
}
\end{figure}

Despite the lack of the tetragonal symmetry, the spatial arrangement of the Cu atoms is close to the square lattice. Therefore, we label individual exchange couplings according to the Cu--Cu distance (see Fig.~\ref{structure}): $J_1,J_1'$, and $J_1''$ run between first (nearest) neighbors, $J_2$ and $J_2'$ run between second (next-nearest) neighbors, etc. The TB fit of the LDA band structure (Fig.~\ref{tb}) identifies the relevant AFM interactions (Table~\ref{tbm}). Surprisingly, the leading AFM interaction is $J_4^{\AFM}\simeq 34$~K (between fourth neighbors) establishing the Cu--Cu dimers that control low-energy magnetic properties, see Fig.~\ref{model}. Other relevant AFM interactions include $J_4'^{\AFM}\simeq 23$~K as well as $J_1^{\AFM}$, $J_2^{\AFM}$, and $J_{\perp}^{\AFM}$ of about $13$~K. Further couplings in the $ab$ plane are below 5~K and can be neglected within a minimum model. The total number of inequivalent exchange couplings up to fourth neighbors in the $ab$ plane amounts to 12. The model approach evaluates all of them, thus simplifying the supercell calculations.

The supercell calculations evaluated: i) all the short-range couplings ($J_1, J_1'$, $J_1''$, $J_2$, and $J_2'$) due to the possible FM contributions; ii) the relevant long-range couplings ($J_4,J_4'$, and $J_{\perp}$). Since all these couplings are relatively weak, they are sensitive to the choice of the $U_{3d}$ parameter in the LSDA+$U$ calculations. Nevertheless, the qualitative scenario is robust with respect to the computational method and can be reproduced for a wide range of $U_{3d}$. In Table~\ref{tbm}, we list the exchange integrals for two representative $U_{3d}$ values that give reasonable agreement with the experimental energy scale, established by the saturation field $\mu_0H_{c2}=30$~T (about 40~K)\cite{kageyama2005-2} and the Curie-Weiss temperature $\theta=10$~K. The different $U_{3d}$ values are required due to the different double-counting correction (DCC) schemes of LSDA+$U$: around-mean-field (AMF), which is the default option in \texttt{FPLO}, and the fully localized limit (FLL) that mimics typical calculations in the \texttt{VASP} code. A similar offset of $3-4$~eV for $U_{3d}$ depending on the DCC has been previously observed in other Cu-containing compounds.\cite{cuncn,dioptase} The generalized gradient approximation (GGA) for the exchange-correlation potential produces nearly the same results. AMF and FLL generally favor AFM and FM couplings, respectively. This can be seen from the Curie-Weiss temperatures, calculated in a mean-field approximation ($\theta=\frac14\sum_i z_iJ_i$, where $z_i$ is the coordination number for $J_i$): $\theta=26$~K for AMF at $U_{3d}=4.5$~eV and $\theta=-17$~K for FLL at $U_{3d}=8.5$~eV.

\begin{table}
\caption{\label{tbm}
Exchange couplings evaluated using the model and supercell approaches (see text). Leading couplings are also shown in Fig.~\ref{structure}. The model approach is based on the hopping parameters $t_i$ that are used to calculate AFM contributions to the exchange $J_i^{\AFM}=4t_i^2/U_{\eff}$ with $U_{\eff}=4$~eV.\cite{cuncn,dioptase} The supercell approach evaluates the total exchange integrals $J_i$ for two different implementations of the LSDA+$U$ method: AMF ($U_{3d}=4.5$~eV) and FLL ($U_{3d}=8.5$~eV), see text for details. 
}
\begin{ruledtabular}
\begin{tabular}{lc@{\hspace{2.5em}}r@{\hspace{1.5em}}r@{\hspace{2.5em}}r@{\hspace{1.5em}}r}
     & Distance & $t_i$   & $J_i^{\AFM}$ & $J_i$, AMF & $J_i$, FLL \\
     & (\r A)   & (meV) & (K) & (K) & (K) \\\hline
$J_1$   & 3.89     &  $-33$  &  13  &  $-43$ &  $-63$ \\
$J_1'$  & 3.64     &  21     &   5  &  $-3$  &  $-3$  \\
$J_1''$ & 4.13     &  20     &   5  &  $-3$  &  $-2$  \\
$J_2$   & 5.43     &  $-35$  &  14  &  33    &  $-6$  \\
$J_2'$  & 5.55     &  $-19$  &   4  &   9    &  $-1$  \\
$J_4$   & 8.65     &  $-54$  &  34  &  54    &  38    \\
$J_4'$  & 8.71     &  $-44$  &  23  &  28    &  14    \\
$J_{\perp}$ & 11.73 & $-35$  &  14  &  16    &  11    \\
\end{tabular}
\end{ruledtabular}
\end{table}

The LSDA+$U$ calculations enable the establishment of the qualitative microscopic scenario. We find sizable AFM interactions $J_4$, $J_4'$, and $J_{\perp}$. The nearest-neighbor coupling $J_1$ is ferromagnetic (FM). In the following, we will use the experimental data to quantify the microscopic model. After the spin lattice and the relevant couplings are established from DFT, efficient numerical techniques evaluate the properties of the respective Heisenberg Hamiltonian and enable the direct comparison to the experiment. Prior to this comparison, we will make additional comments on the structural origin of individual exchange couplings in (CuCl)LaNb$_2$O$_7$.

The FM nature of $J_1$ can be traced back to the twisted configuration of corner-sharing CuO$_2$Cl$_2$ plaquettes. The neighboring plaquettes lie in different planes, thus inducing the very low $J_1^{\AFM}\simeq 13$~K. The weak AFM contribution along with the Hund's coupling on the Cl site\cite{mazurenko2007} lead to the overall FM interaction, despite the Cu--Cl--Cu angle of $107.1^{\circ}$ notably exceeds $90^{\circ}$, where FM superexchange is expected. The leading AFM couplings between the fourth neighbors originate from the effective Cu--Cl--Cl--Cu superexchange pathway with two short Cu--Cl bonds pointing towards each other.\cite{tsirlin2009} Thus, the strong fourth-neighbor coupling is only possible along the $[120]$ and $[1\bar20]$ directions, while the couplings along $[210]$ and $[\bar 210]$ are negligible. The difference between $J_4$ and $J_4'$ is due to the more curved pathway for $J_4'$ (the Cu--Cl--Cl angle of $154^{\circ}$) compared to $J_4$ ($162^{\circ}$). Finally, the sizable coupling $J_{\perp}$ is caused by the low-lying Nb $4d$ states that contribute to the bands near the Fermi level.

The proposed scenario is highly sensitive to the details of the crystal structure. Although relaxation within the DFT framework yields the correct crystal symmetry and the reasonable structural model, fine features of the structure are not properly reproduced. In particular, the Cu--Cl--Cu angle for $J_1$ is overestimated.\cite{tsirlin2009,cheng,refinement} This overestimate makes $J_1$ AFM, while $J_4$ is largely overestimated compared to the experimental energy scale. Thus, the experimental structural information is essential to derive the correct spin model of (CuCl)LaNb$_2$O$_7$.

The resulting spin lattice of (CuCl)LaNb$_2$O$_7$ is shown in Fig.~\ref{model}. Its remarkable feature is the \emph{lack of the magnetic frustration}. The couplings $J_1,J_4$, and $J_4'$ tend to establish columnar AFM ordering with parallel spins along $b$ and antiparallel spins along $a$. Such an ordering is further stabilized by weakly AFM next-nearest-neighbor couplings. To test the proposed spin model against the experimental data and to quantify the exchange couplings, we perform quantum Monte-Carlo (QMC) simulations using the directed loop algorithm in the stochastic series expansion representation, as implemented in the \texttt{ALPS} simulation package.\cite{alps} The typical lattice size was $16\times 16$ (1024 sites, four sites per magnetic unit cell) and allowed to avoid finite-size effects. Magnetic susceptibility and high-field magnetization data were taken from Refs.~\onlinecite{refinement} and~\onlinecite{kageyama2005-2}, respectively.

\begin{figure}
\includegraphics[scale=1]{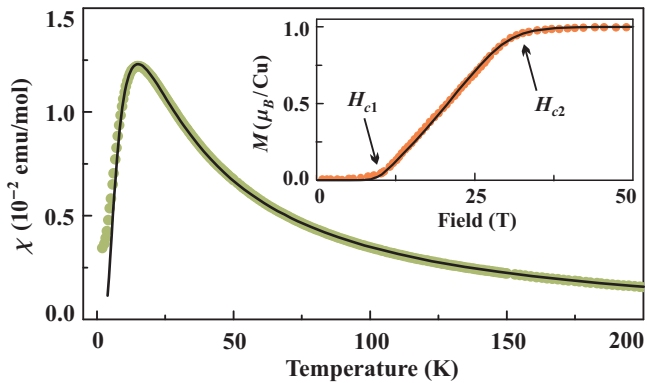}
\caption{\label{experiment}
(Color online) Fits of the experimental data with $J_4=25$~K and $J_2/J_4=0.3$: temperature dependence of the magnetic susceptibility ($\chi$) in the primary figure and field dependence of the magnetization ($M$) in the inset. Experimental data are shown with dots, dark lines are the fits.
}
\end{figure}

Starting from the TB results ($J^{\AFM}$), we first restrict ourselves to the $J_4-J_4'$ alternating chain model and fit the data with $J_4=25$~K and $J_4'/J_4\simeq 0.5$.\cite{note1} However, the energy can be transferred between the bonds of the lattice, leaving some ambiguity for individual $J$'s. For example, we readily obtained another fit with $J_4=25$~K, $J_2/J_4\simeq 0.3$, and $J_4'=0$ (Fig.~\ref{experiment}; the lower $J_2$ is caused by the larger number of the respective bonds).\cite{note2} One can achieve similar fits of the data with an even larger number of parameters, but the individual interdimer couplings remain ambiguous. This implies that the available experimental data are insufficient to evaluate fine details of the (CuCl)LaNb$_2$O$_7$ spin lattice. The fits evidence the intradimer coupling $J_4$ of about 25~K. The interdimer coupling amounts to $50-60$~\% of $J_4$ and can be distributed among different bonds ($J_4',J_2,J_1$, and $J_{\perp}$). To further characterize the spin lattice, inelastic neutron scattering experiments on single crystals are desirable. Presently available powder neutron data\cite{kageyama2005} point to an intradimer distance of 8.8~\r A in remarkable agreement with our model that reveals the dimers on the $J_4$ bond (Cu--Cu distance of 8.65~\r A). 

The proposed spin lattice belongs to the family of coupled spin dimer models. Since the interdimer couplings are non-frustrated, the ground state is determined by the ratio of the intradimer and interdimer couplings. If the intradimer coupling $J_4$ is sufficiently large, the spin gap is opened, as experimentally observed in (CuCl)LaNb$_2$O$_7$. 

The reduced dimerization will close the spin gap and lead to a long-range magnetic ordering. This ordering is of the columnar AFM type, because FM $J_1$ along with AFM $J_2,J_4$, and $J_4'$ stabilize the parallel alignment of spins along the $b$ direction (Fig.~\ref{model}). The columnar AFM ordering has been experimentally observed in the isostructural (CuBr)LaNb$_2$O$_7$ (Ref.~\onlinecite{oba2006}) and (CuCl)LaTa$_2$O$_7$ (Ref.~\onlinecite{kitada2009}). This result demonstrates a broader scope of the proposed spin model. It can be applied to a range of quantum magnets with non-trivial properties. However, the accurate determination of model parameters for (CuBr)LaNb$_2$O$_7$ and (CuCl)LaTa$_2$O$_7$ remains a challenging task and requires the precise structure determination along with the interpretation of the magnetization data. A further challenge is the explanation of the $\frac13$-magnetization plateau in a structurally-related compound (CuBr)Sr$_2$Nb$_3$O$_{10}$.\cite{tsujimoto2007} Such studies are presently underway and will improve our understanding of dimer-based quantum magnets with exotic magnetic behavior.

In summary, we have proposed a valid microscopic model of (CuCl)LaNb$_2$O$_7$. We argue that this compound is a system of spin dimers with non-frustrated interdimer couplings. The intradimer coupling $J_4$ connects fourth neighbors and amounts to 25~K. The interdimer couplings comprise $50-60$~\% of $J_4$ and are distributed among several bonds of the spin lattice. The model is in quantitative agreement with the available experimental data and naturally explains the spin-gap behavior of (CuCl)LaNb$_2$O$_7$ as a result of the dimerization. The limit of the weak dimerization would lead to the columnar antiferromagnetic ordering, relevant for isostructural compounds.

We are grateful to Oleg Janson and Artem Abakumov for supporting discussions. A.Ts. acknowledges the financial support of Alexander von Humboldt Foundation.
\vspace{0.01cm}

\textit{Note added:} After finalizing our manuscript, an independent study of (CuCl)LaNb$_2$O$_7$ appeared.\cite{tassel} Tassel \textit{et al.,} identify (CuCl)LaNb$_2$O$_7$ as a Shatry-Sutherland system with ferromagnetic interdimer couplings. Although the arrangement of dimers resembles our model, we note that the experimental magnetization data could not be fitted with exclusively ferromagnetic interdimer couplings. There are also no experimental indications of the magnetic frustration that is inherent to the Shastry-Sutherland model. Further experimental studies are desirable to resolve the remaining discrepancies. 


\begin{thebibliography}{25}
\bibitem [{not(){\natexlab{a}}}]{note1}%
  \BibitemOpen
  \bibinfo {note} {Further fitting parameters are $\chi_0=-5.6\times
  10^{-4}$~emu/mol and $g=2.15$ in agreement with the Curie-Weiss fit
  ($\chi_0=-5.4\times 10^{-4}$~emu/mol and $\mu_{\eff}=1.87$~$\mu_B$, i.e.,
  $g=2.16$).}\BibitemShut {Stop}%
\end{thebibliography}

%

\newpage
\renewcommand{\thefigure}{S\arabic{figure}}
\setcounter{figure}{0}
\begin{widetext}
\begin{center}
 \large
 \centerline{\textbf{Supplementary material}}
\end{center}
\bigskip

\begin{figure}[!h]
\centerline{\includegraphics{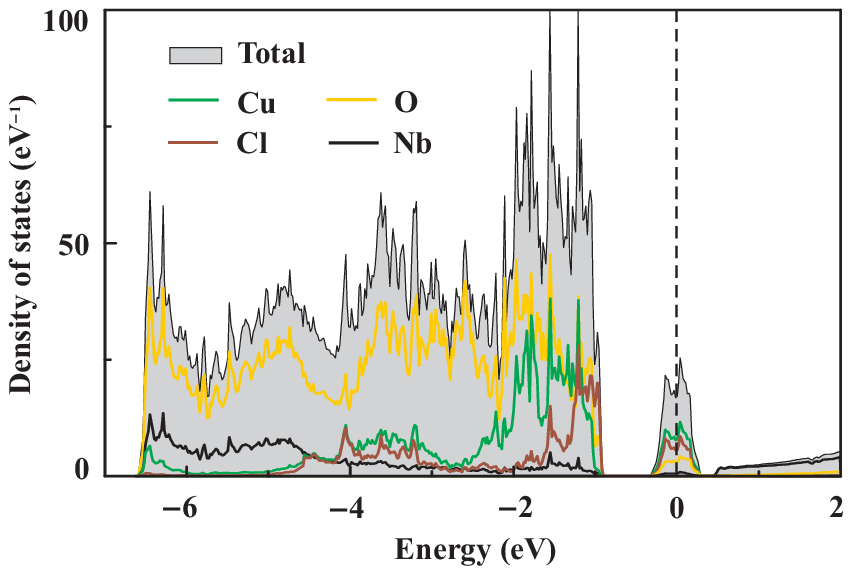}}
\caption{LDA density of states of (CuCl)LaNb$_2$O$_7$. The Fermi level is at zero energy.}
\end{figure}
\medskip

\begin{figure}[!h]
\centerline{\includegraphics{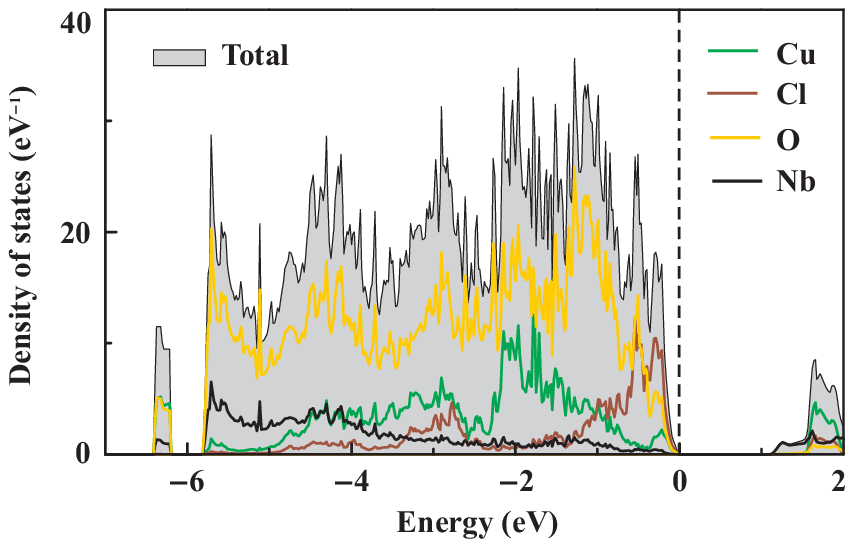}}
\caption{LSDA+$U$ density of states of (CuCl)LaNb$_2$O$_7$ calculated with AMF DCC and $U_{3d}=4.5$~eV for the lowest-energy spin configuration (columnar ordering depicted in Fig.~3 of the paper). The Fermi level is at zero energy.}
\end{figure}
\end{widetext}

\end{document}